\documentclass[twocolumn,showpacs,preprintnumbers,amsmath,amssymb,prl,superscriptaddress]{revtex4}  

\input epsf
\input psfig.sty

\def\ppbar{$p\overline{p}~$}             
\def\met{\mbox{${\hbox{$E$\kern-0.6em\lower-.1ex\hbox{/}}}_T$}} 
\def\mex{\mbox{${\hbox{$E$\kern-0.6em\lower-.1ex\hbox{/}}}_x$}} 
\def\mey{\mbox{${\hbox{$E$\kern-0.6em\lower-.1ex\hbox{/}}}_y$}} 
\def\mexy{\mbox{${\hbox{$E$\kern-0.6em\lower-.1ex\hbox{/}}}_{x,y}$}} 
 
\def\gevcc{GeV/$c^2~$}                   
\def\tevcc{TeV/$c^2~$}                   
\def\gevc{GeV/$c$}                       
\def\gev{GeV}                            
\def\igev{GeV$^{-1}$}                            
\def\D0{D\O}                            
\begin{document}
\lefthyphenmin=2
\righthyphenmin=3
%
%

\title{
Search for Excited and Exotic Muons in the $\mu \gamma$ Decay Channel \\ 
in $\boldmath p\overline{p}$ Collisions at $\boldmath \sqrt{s} = 1.96$ TeV
}


\affiliation{Institute of Physics, Academia Sinica, Taipei, Taiwan 11529, Republic of China} 
\affiliation{Argonne National Laboratory, Argonne, Illinois 60439} 
\affiliation{Institut de Fisica d'Altes Energies, Universitat Autonoma de Barcelona, E-08193, Bellaterra (Barcelona), Spain} 
\affiliation{Baylor University, Waco, Texas  76798} 
\affiliation{Istituto Nazionale di Fisica Nucleare, University of Bologna, I-40127 Bologna, Italy} 
\affiliation{Brandeis University, Waltham, Massachusetts 02254} 
\affiliation{University of California, Davis, Davis, California  95616} 
\affiliation{University of California, Los Angeles, Los Angeles, California  90024} 
\affiliation{University of California, San Diego, La Jolla, California  92093} 
\affiliation{University of California, Santa Barbara, Santa Barbara, California 93106} 
\affiliation{Instituto de Fisica de Cantabria, CSIC-University of Cantabria, 39005 Santander, Spain} 
\affiliation{Carnegie Mellon University, Pittsburgh, PA  15213} 
\affiliation{Enrico Fermi Institute, University of Chicago, Chicago, Illinois 60637} 
\affiliation{Joint Institute for Nuclear Research, RU-141980 Dubna, Russia} 
\affiliation{Duke University, Durham, North Carolina  27708} 
\affiliation{Fermi National Accelerator Laboratory, Batavia, Illinois 60510} 
\affiliation{University of Florida, Gainesville, Florida  32611} 
\affiliation{Laboratori Nazionali di Frascati, Istituto Nazionale di Fisica Nucleare, I-00044 Frascati, Italy} 
\affiliation{University of Geneva, CH-1211 Geneva 4, Switzerland} 
\affiliation{Glasgow University, Glasgow G12 8QQ, United Kingdom} 
\affiliation{Harvard University, Cambridge, Massachusetts 02138} 
\affiliation{Division of High Energy Physics, Department of Physics, University of Helsinki and Helsinki Institute of Physics, FIN-00014, Helsinki, Finland} 
\affiliation{University of Illinois, Urbana, Illinois 61801} 
\affiliation{The Johns Hopkins University, Baltimore, Maryland 21218} 
\affiliation{Institut f\"{u}r Experimentelle Kernphysik, Universit\"{a}t Karlsruhe, 76128 Karlsruhe, Germany} 
\affiliation{High Energy Accelerator Research Organization (KEK), Tsukuba, Ibaraki 305, Japan} 
\affiliation{Center for High Energy Physics: Kyungpook National University, Taegu 702-701, Korea; Seoul National University, Seoul 151-742, Korea; and SungKyunKwan University, Suwon 440-746, Korea} 
\affiliation{Ernest Orlando Lawrence Berkeley National Laboratory, Berkeley, California 94720} 
\affiliation{University of Liverpool, Liverpool L69 7ZE, United Kingdom} 
\affiliation{University College London, London WC1E 6BT, United Kingdom} 
\affiliation{Centro de Investigaciones Energeticas Medioambientales y Tecnologicas, E-28040 Madrid, Spain} 
\affiliation{Massachusetts Institute of Technology, Cambridge, Massachusetts  02139} 
\affiliation{Institute of Particle Physics: McGill University, Montr\'{e}al, Canada H3A~2T8; and University of Toronto, Toronto, Canada M5S~1A7} 
\affiliation{University of Michigan, Ann Arbor, Michigan 48109} 
\affiliation{Michigan State University, East Lansing, Michigan  48824} 
\affiliation{Institution for Theoretical and Experimental Physics, ITEP, Moscow 117259, Russia} 
\affiliation{University of New Mexico, Albuquerque, New Mexico 87131} 
\affiliation{Northwestern University, Evanston, Illinois  60208} 
\affiliation{The Ohio State University, Columbus, Ohio  43210} 
\affiliation{Okayama University, Okayama 700-8530, Japan} 
\affiliation{Osaka City University, Osaka 588, Japan} 
\affiliation{University of Oxford, Oxford OX1 3RH, United Kingdom} 
\affiliation{University of Padova, Istituto Nazionale di Fisica Nucleare, Sezione di Padova-Trento, I-35131 Padova, Italy} 
\affiliation{LPNHE, Universite Pierre et Marie Curie/IN2P3-CNRS, UMR7585, Paris, F-75252 France} 
\affiliation{University of Pennsylvania, Philadelphia, Pennsylvania 19104} 
\affiliation{Istituto Nazionale di Fisica Nucleare Pisa, Universities of Pisa, Siena and Scuola Normale Superiore, I-56127 Pisa, Italy} 
\affiliation{University of Pittsburgh, Pittsburgh, Pennsylvania 15260} 
\affiliation{Purdue University, West Lafayette, Indiana 47907} 
\affiliation{University of Rochester, Rochester, New York 14627} 
\affiliation{The Rockefeller University, New York, New York 10021} 
\affiliation{Istituto Nazionale di Fisica Nucleare, Sezione di Roma 1, University of Rome ``La Sapienza," I-00185 Roma, Italy} 
\affiliation{Rutgers University, Piscataway, New Jersey 08855} 
\affiliation{Texas A\&M University, College Station, Texas 77843} 
\affiliation{Istituto Nazionale di Fisica Nucleare, University of Trieste/\ Udine, Italy} 
\affiliation{University of Tsukuba, Tsukuba, Ibaraki 305, Japan} 
\affiliation{Tufts University, Medford, Massachusetts 02155} 
\affiliation{Waseda University, Tokyo 169, Japan} 
\affiliation{Wayne State University, Detroit, Michigan  48201} 
\affiliation{University of Wisconsin, Madison, Wisconsin 53706} 
\affiliation{Yale University, New Haven, Connecticut 06520} 
\author{A.~Abulencia}
\affiliation{University of Illinois, Urbana, Illinois 61801}
\author{D.~Acosta}
\affiliation{University of Florida, Gainesville, Florida  32611}
\author{J.~Adelman}
\affiliation{Enrico Fermi Institute, University of Chicago, Chicago, Illinois 60637}
\author{T.~Affolder}
\affiliation{University of California, Santa Barbara, Santa Barbara, California 93106}
\author{T.~Akimoto}
\affiliation{University of Tsukuba, Tsukuba, Ibaraki 305, Japan}
\author{M.G.~Albrow}
\affiliation{Fermi National Accelerator Laboratory, Batavia, Illinois 60510}
\author{D.~Ambrose}
\affiliation{Fermi National Accelerator Laboratory, Batavia, Illinois 60510}
\author{S.~Amerio}
\affiliation{University of Padova, Istituto Nazionale di Fisica Nucleare, Sezione di Padova-Trento, I-35131 Padova, Italy}
\author{D.~Amidei}
\affiliation{University of Michigan, Ann Arbor, Michigan 48109}
\author{A.~Anastassov}
\affiliation{Rutgers University, Piscataway, New Jersey 08855}
% ...
\collaboration{CDF Collaboration}
\noaffiliation

%
\begin{abstract}

We present a search for excited and exotic muon states $\mu^*$, conducted using 
 an integrated luminosity of 371 pb$^{-1}$ of data collected in 
 {\mbox{$p\bar p$}}\ collisions at {\mbox{$\sqrt{s}$ =\ 1.96\ TeV }}
  at the Tevatron with the CDF II detector.  We search for associated 
 production of $\mu \mu^*$ followed by the decay 
$\mu^* \rightarrow \mu \gamma$, 
 resulting in the $\mu \mu \gamma$ final state.  
 We compare the data to model predictions as a function of the mass of the 
 excited muon $M_{\mu^*}$, the compositeness energy scale $\Lambda$, 
 and the gauge coupling factor $f$.  
 No signal above the 
 standard model expectation is observed in the $\mu \gamma$ mass spectrum. 
 In the contact interaction model, we exclude 
 $107 < M_{\mu^*} < 853$ \gevcc for $\Lambda = M_{\mu^*}$;
 in the gauge-mediated model, we exclude $100 < M_{\mu^*} < 410$ \gevcc for 
 $f/\Lambda  = 10^{-2}$~\igev.  These 95\% confidence level
 exclusions extend previous limits and are the first hadron collider 
 results on $\mu^*$ production in the gauge-mediated model.
\end{abstract}


\maketitle


 In the standard model (SM) the quarks
 and leptons are treated as fundamental particles. Their 
 generational structure and observed mass hierarchy motivate a model of 
 composite quarks and leptons consisting of fewer elementary 
 particles than contained in the SM~\cite{baurestar}.
 In this model, quarks and leptons are the
 lowest-energy bound states of these hypothetical particles, and
 additional excited states exist near the compositeness energy
 scale $\Lambda$.

 Exotic fermions are also predicted in the context
 of grand unified or string theories, in which the known forces
 are unified into a larger symmetry group~\cite{gmestar}.  In such models,
 additional fermions are predicted with properties similar to those of
 excited fermions.

 At a \ppbar collider, excited or exotic muon states could be
 observed through the reaction $q \bar{q} \rightarrow \mu^* \mu$.  
 Excited muon production can be described using a contact interaction 
 (CI) Lagrangian density~\cite{baurestar}:
\begin{equation}
L = \frac{4\pi}{\Lambda^2}\bar q_L \gamma^\mu q_L \bar  M_L \gamma_\mu \mu_L + h.c., \quad \nonumber
\end{equation} 
 where $M_L$ represents the left-handed $\mu^*$ field.  For exotic muon
production, the relevant gauge-mediated (GM) Lagrangian density is~\cite{gmestar}:
\begin{equation}
L = \frac{1}{2\Lambda} \bar M_R \sigma^{\mu \nu} \left[ fg \frac{\vec \tau}{2} \cdot {\vec W}_{\mu \nu} 
+ f'g' \frac{Y}{2} B_{\mu \nu} \right] \mu_L + h.c., \nonumber
\end{equation} 
 where $W$ and $B$ are the $SU(2)_L$ and $U(1)_Y$ field strengths, 
 $g$ and $g^\prime$ are the respective electroweak couplings, and 
 $f$ and $f^\prime$ are phenomenological
 constants which are set equal to each other by convention to 
 maximize the photonic decay branching ratio.  
 For $f = f^\prime$ ($f = -f^\prime$),
the relative $\mu^*$ branching ratios are BR$(\mu^* \rightarrow \mu \gamma) 
\approx 0.3$ ($0$), BR$(\mu^* \rightarrow \nu W) 
\approx 0.6$ ($0.6$), and BR$(\mu^* \rightarrow \mu Z) 
\approx 0.1$ ($0.4$) for $M_{\mu^*} > 200$ GeV/$c^2$.  
 The BR$(\mu^* \rightarrow \mu \gamma)$ increases to 70\% at
 $M_{\mu^*} = 100$ GeV/$c^2$ for $f = f^\prime$. We use these branching 
 ratios for both the GM and CI production models~\cite{cdfestar}.

 In this Letter we describe a search for associated $\mu \mu^*$ production 
 that extends existing $\mu^*$ mass limits. 
 Prior searches for $\mu^*$ production have been performed by the LEP 
 experiments, which have excluded $\mu^*$ with $M_{\mu^*} < 200 $ 
 \gevcc\ for $f/\Lambda > 10^{-2}$~\igev\ in the GM model~\cite{opal}.
 The D\O\ experiment has excluded $\mu^*$ with 
 $M_{\mu^*} < 688$ \gevcc\ for $\Lambda = M_{\mu^*}$ in
 a particular CI model~\cite{d0}.

 We use 371 pb$^{-1}$ of data collected with the CDF II detector at the 
 Fermilab Tevatron. The detector has been described in detail elsewhere~\cite{tdr}. 
 Its magnetic spectrometer consists of silicon microstrip and drift 
 chamber~\cite{cot} tracking detectors.  Surrounding this are 
 central~\cite{run1detector} ($|\eta|<1$~\cite{perpendicular}) and forward ~\cite{plugcal} ($|\eta|>1.1$) 
 electromagnetic (EM) and hadronic calorimeters.
 Embedded in the central EM calorimeter are wire and strip chambers~\cite{tdr}
 (used to measure the transverse shower profiles of photons) and 
 a central preshower detector~\cite{tdr} (used for detecting photon
 conversions).
 Outside the calorimeters are CMUP ($|\eta|<0.6$) and CMX
 ($0.6<|\eta|<1$) muon detectors~\cite{muonsystem}. 
 The momentum resolution of beam-constrained drift-chamber tracks is 
 $\delta p_T/p_T^2 \approx 0.05$\%$/$(\gevc).  The electromagnetic energy 
 resolution for the high-energy photons typical of $\mu^*$ decays is approximately 2.5\%.

 We analyze events passing the readout trigger requirement of one drift-chamber 
 track with $p_T> 18$~\gevc~\cite{perpendicular} matched to a reconstructed track 
 segment in the muon chambers.
 In the offline analysis, we require two muon candidates identified by
 drift-chamber tracks with $p_T>20$~\gevc\ and $|\eta|<1$, which pass requirements on 
 impact parameter and number of hits, have minimum-ionizing particle properties, and 
 at least one of which has a matching muon chamber segment~\cite{wzprd}.  Both tracks must pass 
 the calorimeter isolation requirement $I < 0.1$, where $I$ is the 
 ratio of the combined EM and hadronic calorimeter $E_T$ surrounding the track in a 
 cone of radius $R = \sqrt{(\Delta \eta)^2 + (\Delta \phi)^2} < 0.4$~\cite{perpendicular} to the 
 track $p_T$. 
 In addition, both tracks must also pass the 
 tracking isolation requirement $T < 0.1$, where $T$ is the ratio of the summed 
 $p_T$ of tracks to the muon $p_T$, where the sum is taken over tracks in a cone of 
 radius $R<0.4$ around the muon direction.  Finally, we reject cosmic
 rays based on tracking and track-timing information~\cite{cotcosmics}.

 We select dimuon events that have a central or forward photon 
 candidate with $E_T > 25$ \gev, $|\eta|<2.8$, and $I<0.1$, where 
 calorimeter isolation is now defined relative to 
 the photon.  The photon energy is defined as the EM calorimeter energy
 in a $0.2 \times 0.25\ \eta - \phi$ region, and its transverse component
 is determined by the vertex of the highest momentum muon with a 
 matching muon chamber segment.
 Photon candidates are identified on the basis  
 of their longitudinal and transverse calorimeter shower profiles. 
 Central photon candidates require the sum $p_T$ of charged particles
 within a cone of radius $R<0.4$ around the photon direction to be less than
 $5~{\rm GeV} + 0.005 \times E_T^\gamma$, where $E_T^\gamma$ is the $E_T$ of the photon,
 and no single charged particle within $R<0.4$ to have  
 $p_T > ( 1 + 0.005 \times E_T^\gamma/{\rm GeV})$ \gevc.
 To suppress the initial-state radiation (ISR) 
 $Z(\to \mu\mu)\gamma$ background, we reject events with 
 dimuon invariant mass $m_{\mu \mu}$ in the range $81 - 101$ GeV/$c^2$.

 A $Z \rightarrow \mu \mu$ sample is used to measure the efficiencies of the muon 
 identification criteria and trigger.
 The efficiency of the calorimeter and tracking identification requirements 
 is measured to be $(92.6 \pm 0.3_{stat})$\%.  We 
 measure the combined trigger and muon chamber matching efficiency to be 
 $(79.3 \pm 1.0_{stat})$\%  for the CMUP and $(95.0 \pm 0.6_{stat})$\% 
 for the CMX.
 
 The photon identification efficiency is extracted from a 
 {\sc geant}-based detector simulation~\cite{geant}.  Since
 the electromagnetic showers of electrons are similar to those of photons, 
 we validate the simulated photon efficiency using a control sample of electrons 
 from $Z \rightarrow ee$ events. 
 The events are triggered and identified using one electron; the second electron
 is used to emulate the photon to determine the cut efficiencies for
 photons.
 In the central calorimeter, the emulated photon efficiency is 
 $(79.0 \pm 0.5_{stat})$\% in the data
 and $(77.0 \pm 0.6_{stat})$\% in the simulation.
 The corresponding values in the forward calorimeter are
 $(89.0 \pm 0.6_{stat})$\% in the data and $(89.6 \pm 0.6_{stat})$\%
 in the simulation.  These comparisons
 show that photon detection is well-modeled by the simulation; the statistically
 significant differences are included in the systematic uncertainty.

 The geometric acceptance and detector response are calculated with the {\sc geant} 
 detector simulation separately for the CI and GM models. We use the 
 {\sc pythia}~\cite{pythia} generator for the CI model and the 
 {\sc lanhep}~\cite{lanhep} and {\sc comphep}~\cite{comphep}
 programs to generate GM model events. The 
 total signal acceptance (including identification efficiencies) for 
 the CI model increases from 
 13\% at $M_{\mu^*} = 100$~\gevcc\ to an asymptotic value of
 21\% for $M_{\mu^*} > 400$~\gevcc.  
 The acceptance for the GM  model increases from 
 12\% at $M_{\mu^*} = 100$~\gevcc\ to an asymptotic value of 
 23\% for $M_{\mu^*} > 300$~\gevcc.  
 The relative systematic uncertainty on 
 the acceptance is 3.1\%, which is dominated by the uncertainty 
 in the identification efficiency (2.2\%) and simulation statistics (2.0\%).

 We compute the expected background contributions from the following sources: 
 (1) $Z/\gamma^* ~(+ \gamma) \rightarrow \mu \mu \gamma$, 
 (2) $Z/\gamma^* ~(+ \gamma) \rightarrow \tau \tau \gamma$, where the $\tau$'s decay to muons, 
 (3) $Z/\gamma^* ( \rightarrow \mu \mu ) +$ jet, where the jet is misidentified as a photon, 
 (4) $t (\rightarrow \mu \nu b) + \bar{t}(\rightarrow \mu \nu \bar{b})$, where a 
  fermion radiates a photon,
 (5) $W ( \rightarrow e \nu ) + Z (\rightarrow \mu \mu)$, where the electron is 
 misidentified as a photon, and 
 (6) $ Z ( \rightarrow e e ) + Z ( \rightarrow \mu \mu)$, where one of the electrons 
 is misidentified as a photon. 
 Other backgrounds ($\geq 3$ jets, $W + \geq 2$ jets, and 
 $W\gamma + \geq 1$ jet) were found to be negligible.  

The $Z\gamma$, $t \bar{t}$, $WZ$ and $ZZ$ backgrounds 
 are estimated 
 using simulated events, with the {\sc zgamma}~\cite{zgamma} generator 
 for the $Z (\rightarrow \mu \mu) \gamma$ 
 background and {\sc pythia} for the others.  Systematic uncertainties
 on these background predictions arise due to integrated luminosity 
 (6\%)~\cite{luminosity}, parton distribution functions (PDFs) (5\%), 
 higher-order QCD corrections (5\%)~\cite{vanneerven}, acceptance (1\%), 
 and identification efficiencies (2\%). 

 The $Z+$jet background is estimated by weighting $Z+$jet events from 
 the data by an $E_T$-dependent jet~$\rightarrow \gamma$ misidentification rate. 
 The latter is measured using a jet-triggered data sample, 
 correcting for the fraction of true prompt photons in the jet sample~\cite{cdfestar}.
 The prompt photon fraction is estimated using $\gamma\rightarrow ee$ 
 conversions identified with the calorimeter preshower 
 detector~\cite{hkgthesis}. The jet$\rightarrow \gamma$ misidentification rate
 is applied as a function of $E_T$ in the central calorimeter and as a 
 function of $E_T$ and $\eta$ in the forward calorimeter. 

\begin{figure}[!htbp]
\begin{center}
\epsfysize = 4.cm
\epsffile{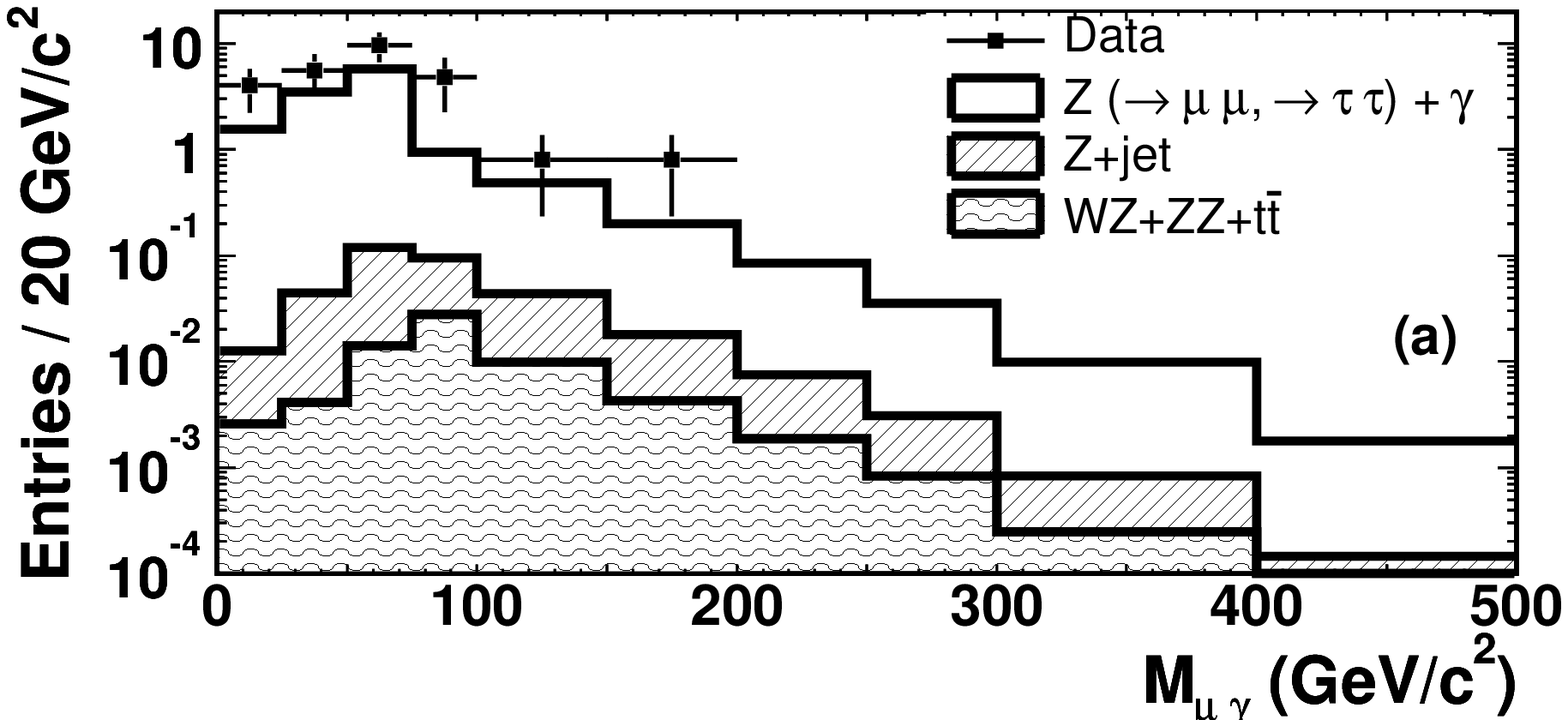}
\vspace*{1.5mm}
\hspace*{1.5mm}
\epsfysize = 4.cm
\epsffile{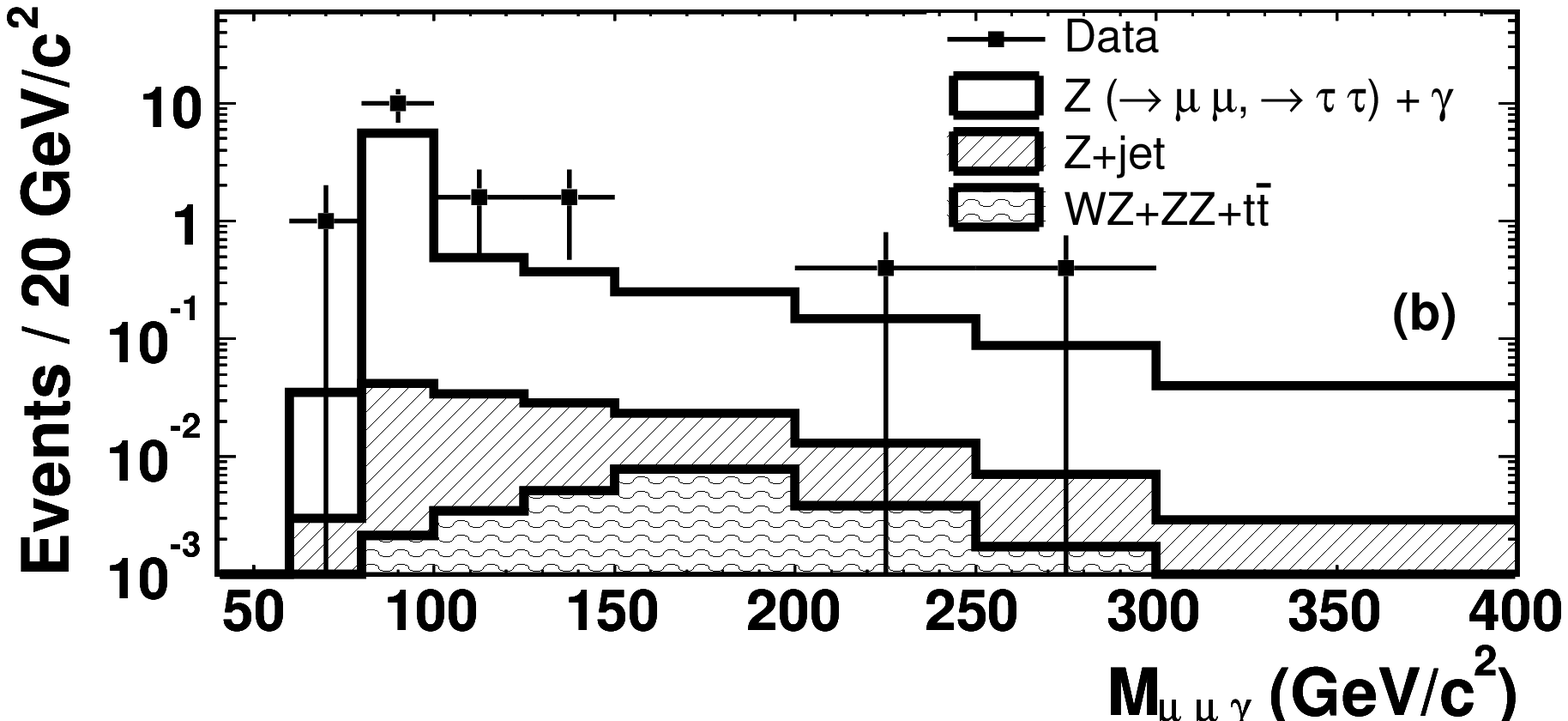}
\vspace*{1mm}
\caption{The $\mu\gamma$ (a) and $\mu\mu\gamma$ (b) mass distributions for 
 the data with background expectations.  The total number of observed 
 (expected) $\mu\mu\gamma$ entries is 17 ($8.3 \pm 0.9$).  Both $\mu\gamma$
 combinations per event are included in (a). \label{backgrounds}}
\end{center}
\end{figure}

  We observe 17 signal candidates with a background prediction of 
 $8.3 \pm 0.9$, of which $8.1 \pm 0.8$ are expected from $Z\gamma$
 production.  
 The Poisson probability for the 
 background to fluctuate up to the observation, or higher, is 0.8\%.
 Investigating further, we find that 11 candidate events have
 a 3-body mass $m_{\mu \mu \gamma}$ in the 81-101~\gevcc\ range, 
 consistent with being final-state radiation (FSR) $Z \to \mu \mu \gamma$ events, 
 to be compared to the FSR $Z \to \mu \mu \gamma$ prediction of $5.5 \pm 0.5$ events.
 Figure~\ref{backgrounds} and Table~\ref{tab:IntegratedBg} 
 show the $m_{\mu \gamma}$ and $m_{\mu \mu \gamma}$ distributions for the
 data and background. 

\begin{figure}[!htbp]
\begin{center}
\epsfysize = 4.cm
\epsffile{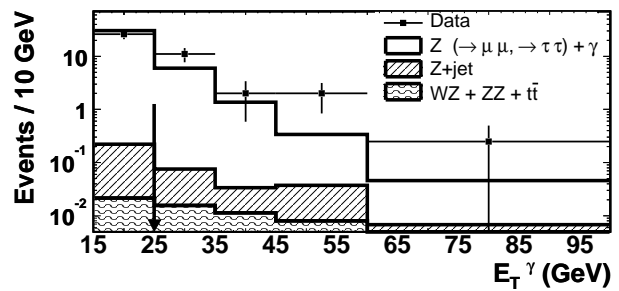}
\vspace*{1mm}
\caption{The $E_{T}^\gamma$ distributions for the 
 data (43 events) and background ($38.5 \pm 4.0$ events) 
 expectations for $E_{T}^\gamma> 15\ {\rm GeV}$.  The arrow indicates
 the minimum $E_{T}^\gamma$ required for this analysis.}
\label{phoetbackgrounds}
\end{center}
\end{figure}

 To test our background prediction,
 we lower the minimum photon $E_T$ to 15 GeV and observe 43 events with 
 a prediction of $38.5 \pm 4.0$ events.  The data show 
 good agreement with expectation in this higher statistics sample, as shown 
 in Figure~\ref{phoetbackgrounds}. 
 Additionally, we find consistency in the ISR 
 $Z \gamma$ control region of $81 < m_{\mu \mu} < 101$~\gevcc\
 and $25 < E_T^\gamma < 50$ GeV, where we observe 5 events with a
 prediction of $7.2 ^{+1.2}_{-0.8}$ events.
 We conclude that our signal sample has an upward statistical fluctuation, 
 dominantly in the number of FSR $Z \to \mu \mu \gamma$  events.  
 
\begin{table}
\caption{\label{runningSum} Comparison of data and
 integrated background predictions
  above a given cut on the invariant mass of 
 all $\mu\gamma$ combinations (left) and on the $\mu\mu\gamma$ 
 invariant mass (right).}
\begin{tabular}{ccc|ccc}
\hline
\hline
\multicolumn{3}{c|}{$\mu\gamma$ combinations}&\multicolumn{3}{c}{Events}\\
$m_{\mu\gamma}$  & Data & Background & $m_{\mu\mu\gamma}$  & Data & Background \\
 (\gevcc)      &   &                    & (\gevcc) &   &  \\
\hline
$>0$    & 34& $16.6 \pm 1.8 $    & $>0$    &17 & $8.3 \pm 0.9       $\\ 
$>50$   & 22& $10.4 \pm 1.1 $    & $>100$  & 6 & $2.7 \pm 0.3       $\\
$>100$  & 4 & $2.1 \pm 0.3 $     & $>150$  & 2 & $1.5 \pm 0.2 $\\
$>150$  & 2 & $0.89 \pm 0.14$    & $>200$  & 2 & $0.9 \pm 0.1 $ \\
$>200$  & 0 & $0.37 \pm 0.07$    & $>250$  & 1 & $0.51 \pm 0.09$ \\
       &   &                    & $>300$  & 0 & $0.29 \pm 0.06$ \\
\hline
\hline
\end{tabular}
\label{tab:IntegratedBg}
\end{table}
    
 For the $\mu^*$ resonance search, we scan the $m_{\mu \gamma}$ spectrum 
 with a sliding window of width $3 \sigma$, where $\sigma$ is the mass 
 width predicted by the simulation.  Over almost the 
 entire model parameter space, $\sigma$ is dominated by detector resolution.  
 The tracker momentum scale and resolution, and the calorimeter energy scale 
 and resolution, are tuned on the well-known $Z \rightarrow \mu \mu$ and
 $Z \rightarrow e e$ mass peaks~\cite{zmass}, respectively.
 For $M_{\mu^*}=\Lambda$, the reconstructed $\mu\gamma$ mass resolution 
 ranges from 9-90 \gevcc for masses
 ranging from 200-800 GeV/$c^2$.

 We use a Bayesian~\cite{bayes} approach, with a flat prior on the
 signal cross section and gamma priors on acceptance and backgrounds,
 to set limits on the $\mu \mu^*$ production 
 cross section as a function of $M_{\mu^*}$~\cite{limitevent}.
  The cross section limits are converted to 
 $M_\mu^*$ limits by comparing them to the next-to-next-to-leading-order (NNLO) 
theoretical cross sections \cite{vanneerven}, 
 computed using the MRST set of PDFs \cite{mrst}.  We use the CTEQ prescription 
\cite{cteq} to calculate the cross section uncertainty due to PDFs, which varies from 
5\% ($M_{\mu^*} = 100$ \gevcc) to 20\% ($M_{\mu^*} = 1$ \tevcc).  Uncertainties on higher-order 
QCD corrections (7-13\%) depend on $M_{\mu^*}$ and the production model.

\begin{figure}[!htbp]
\begin{center}
\epsfysize = 4.4cm
\epsffile{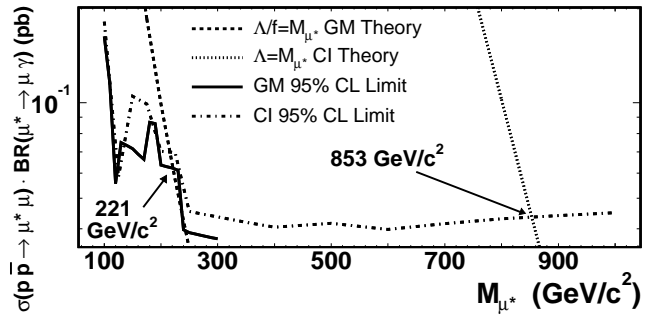}
\vspace*{1mm}
\caption{The experimental cross section $\times$ branching ratio 
limits at 95\% CL for the CI (dashed-dotted line) and GM models (solid line),
compared to the CI model 
 prediction for $ \Lambda = M_{\mu^*}$ (dotted line)
  and the GM model prediction for $ \Lambda / f
 = M_{\mu^*}$ (dashed line).  Also indicated are the mass values 
 that are excluded by these data.}
\label{crossSectionLimit}
\end{center}
\end{figure}

\begin{figure}[!htbp]
\begin{center}
\vspace*{5mm}
\epsfysize = 4.cm
\epsffile{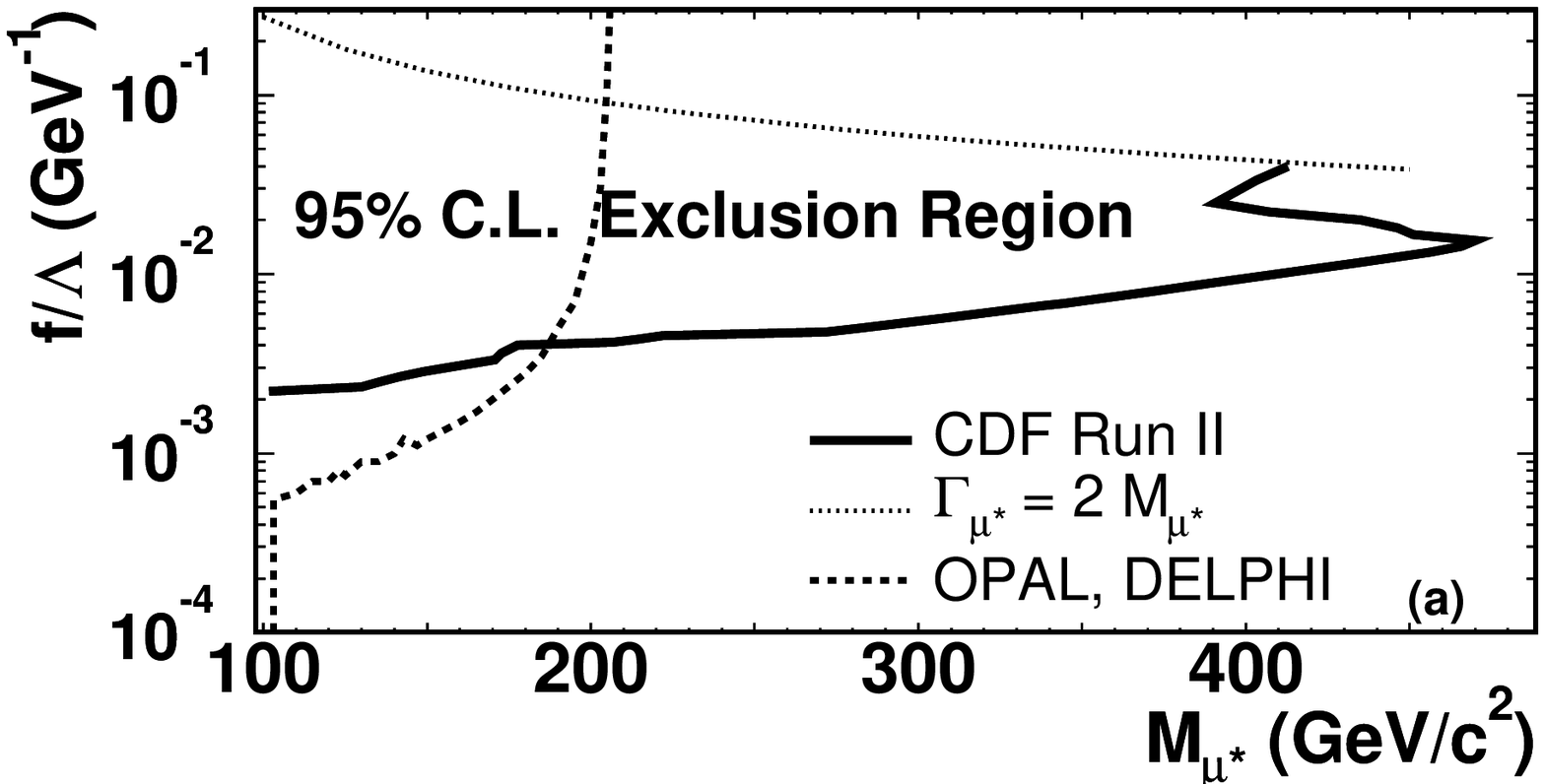}
\vspace*{1mm}
\epsfysize = 4.cm
\epsffile{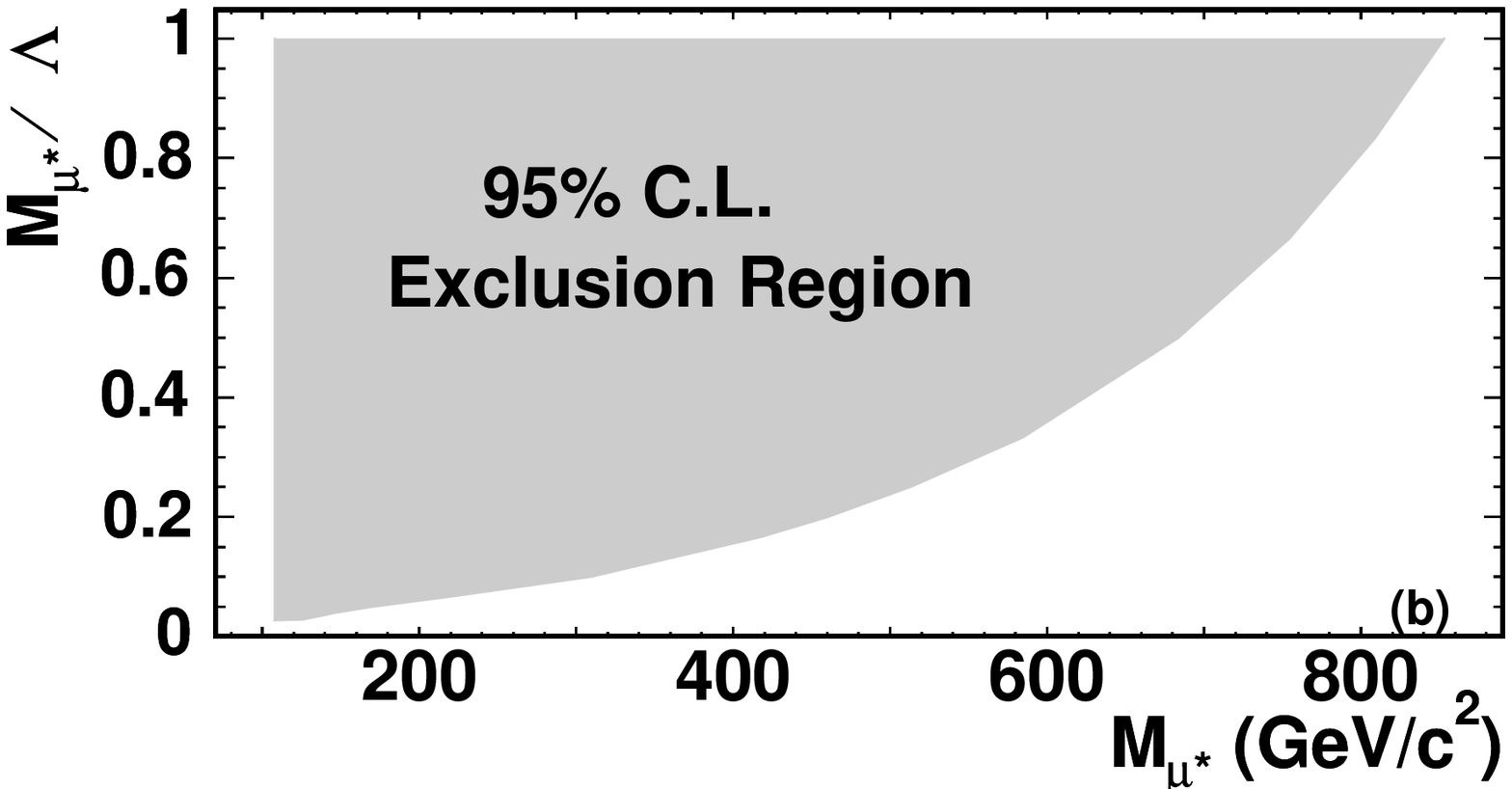}
\vspace*{1mm}
\caption{The 2-D parameter space regions  excluded by this analysis for
 (a) the GM model, along with the current world limits~\cite{opal}, 
and (b) the CI model.  We do not consider the region 
$\Gamma_{\mu^*} > 2 M_{\mu^*}$ where the total width is larger than 
the $\mu^*$ mass. }
\label{limits}
\end{center}
\end{figure}

  Figure~\ref{crossSectionLimit} shows the 95\% confidence level (CL)
 upper limits on the experimental
 cross section $\times$ branching ratio and the theoretical curves 
 for $M_{\mu^*} = \Lambda/f$ ($M_{\mu^*} = \Lambda$)
 for the GM (CI) model.  Masses below
 221 \gevcc (853 GeV/$c^2$) are excluded for the GM (CI) model for the case
 $M_{\mu^*} = \Lambda/f$ ($M_{\mu^*} = \Lambda$). 
  Figure~\ref{limits}
 shows the limits in the parameter space of $f / \Lambda$ ($M_{\mu^*} / \Lambda$)
 versus $M_{\mu^*}$ for the GM (CI) model.  For both models, we lose sensitivity
 for $M_{\mu^*} < 100$ \gevcc due to large backgrounds and loss of signal acceptance
 in this range.  We note that our CI limit of $M_{\mu^*} > 853$ \gevcc\
 reduces to $M_{\mu^*} > 696$ \gevcc\ if we use $\mu^*$ branching ratios
 that account for hypothetical CI decays, as
 assumed by the D\O\ collaboration \cite{d0}.

In conclusion, we have presented a search for
 excited and exotic muons. 
 In the GM model, we exclude  $100 < M_{\mu^*} < 410$ \gevcc for 
 $f/\Lambda = 10^{-2}$~\igev\ 
 at the 95\% C.L., which is the first GM result at a hadron collider
 and is substantially beyond previous limits~\cite{opal}.
 We have also presented $\mu^*$ limits in the
 CI model as a function of $M_{\mu^*}$ and $\Lambda$, excluding
 $107 < M_{\mu^*} < 853$ \gevcc\ for $ \Lambda = M_{\mu^*}$.
These limits complement our recent results of a search for excited and 
exotic electrons \cite{cdfestar}.

We are grateful to Alejandro Daleo for providing NNLO cross section
 calculations. 
We thank the Fermilab staff and the technical staffs of the participating institutions for their vital contributions. This work was supported by the U.S. Department of Energy and National Science Foundation; the Italian Istituto Nazionale di Fisica Nucleare; the Ministry of Education, Culture, Sports, Science and Technology of Japan; the Natural Sciences and Engineering Research Council of Canada; the National Science Council of the Republic of China; the Swiss National Science Foundation; the A.P. Sloan Foundation; the Bundesministerium f\"ur Bildung und Forschung, Germany; the Korean Science and Engineering Foundation and the Korean Research Foundation; the Particle Physics and Astronomy Research Council and the Royal Society, UK; the Russian Foundation for Basic Research; the Comisi\'on Interministerial de Ciencia y Tecnolog\'{\i}a, Spain; in part by the European Community's Human Potential Programme under contract HPRN-CT-2002-00292; and the Academy of Finland. 
%



\begin{thebibliography}{99}
\bibitem{perpendicular}  
 CDF uses a cylindrical coordinate system in which $\phi$ is the azimuthal angle,
 $r$ is the radius from the nominal beamline, and $+z$ points in the direction of the 
 proton beam and is zero at the center of the detector.  The pseudorapidity is defined as
 $\eta = -$ln[$\tan(\theta/2)$], where $\theta$ is the polar angle with respect
 to the $z$ axis.  Calorimeter energy (track momentum) measured transverse to
 the beam is denoted as $E_T$ ($p_T$).
\bibitem{limitevent}  
 An event is considered a $\mu^*$ candidate for a given mass hypothesis if at 
 least one $\mu\gamma$ combination has a reconstructed mass in the sliding mass window.
\end{thebibliography}
\end{document}